\documentclass[apj]{emulateapj}

\usepackage{times}

\shorttitle{Inverse FIP effect in solar flares}                    
\shortauthors{Brooks}

\begin{document}


\title{A diagnostic of coronal elemental behavior during the inverse FIP effect in solar flares}

\author{David H. Brooks\altaffilmark{1,2}}

\affiliation{\altaffilmark{1}College of Science, George Mason University, 4400 University Drive,
  Fairfax, VA 22030}

\altaffiltext{2}{Current address: Hinode Team, ISAS/JAXA, 3-1-1 Yoshinodai, Chuo-ku, Sagamihara,
  Kanagawa 252-5210, Japan}


\begin{abstract}

The solar corona shows a distinctive pattern of elemental abundances that is different from
that of the photosphere. Low first ionization potential (FIP) elements are enhanced by factors
of several. A similar effect is seen in the atmospheres of some solar-like stars, while late
type M stars show an inverse FIP effect. This inverse effect was recently detected on the Sun
during solar flares, potentially allowing a very detailed look at the spatial and temporal
behavior that is not possible from stellar observations. A key question for interpreting these 
measurements is whether both effects act solely on low FIP elements (a true inverse effect
predicted by some models), or whether the inverse FIP effect arises because high FIP elements are enhanced. Here we develop a new
diagnostic that can discriminate between the two scenarios, 
based on modeling of the radiated
power loss, and applying the models to a numerical hydrodynamic simulation of coronal loop
cooling. 
We show that when low/high FIP elements are 
depleted/enhanced, there is a significant difference in the cooling lifetime of loops that is
greatest at lower temperatures. We apply this diagnostic to a post X1.8 flare loop arcade
and inverse FIP region, and show that for this event, low FIP elements are depleted. We 
discuss the results in the context of stellar observations, and models of the FIP and 
inverse FIP effect. We also provide the radiated power loss functions for the two inverse FIP
effect scenarios in machine readable form to facilitate further modeling.
\end{abstract}


 
\section{introduction}

Despite the early UV and X-ray spectroscopic work of \citet{pottasch_1963}, it was not until the mid-1980s
that it was recognized that
the elemental composition of the solar photosphere is different than that of the corona and slow speed
solar wind \citep{meyer_1985}. Elements with a low
first ionization potential (FIP), below around $\sim$ 10\,eV, are enhanced in the corona and slow wind
by factors
of 2--4 compared to their photospheric values \citep{feldman_1992}. This abundance anomaly is known as the FIP effect and is
likely related to the coronal heating mechanism itself. 
A large body of work also now exists exploring coronal abundance anomalies in evolving active regions and 
different features within them 
\citep{sheeley_1995, widing_1997, widing&feldman_2001, testa_etal2011, warren_etal2012, baker_etal2013, baker_etal2015, delzanna_2013b, delzanna&mason_2014}, 
and also as an identification diagnostic for sources of the solar wind
\citep{ko_etal2006, brooks&warren_2011, brooks&warren_2012, brooks_etal2015, lee_etal2015, guennou_etal2015}. 
\citet{schmelz_etal2012} gives a recent review of our
state of knowledge of solar coronal abundances.

The effect also appears to be present in stellar coronae \citep{drake_etal1997}; and see \citet{feldman&laming_2000},
\citet{testa_2010}, and \citet{testa_etal2015} for reviews. In fact, stellar coronae show an interesting pattern.
Stars of a similar spectral type to the Sun show a solar-like FIP effect, transitioning through no obvious effect
around spectral type K5, to an inverse FIP effect in M-type stars \citep{wood&linsky_2010, wood_etal2012}. As pointed out
by \citet{laming_2015}, stars of later spectral type likely have a greater preponderance of large starspots with strong
umbral and penumbral field. So greater flaring activity, for example, could be consistent with the absence of the
FIP effect in early skeptical work on abundance variations in solar flares \citep{feldman&widing_1990, phillips_etal1994},
that are now potentially understood as consistent with the idea that the FIP effect does not operate when material
is rapidly ejected from below the photosphere as in flares \citep{warren_2014}, impulsive heating events \citep{warren_etal2016}, 
and coronal jets \citep{lee_etal2015}; but see \citet{sterling_etal2015}).

Solar and stellar observations have complimentary advantages that we can use to move towards the goal of understanding
how their atmospheres are heated, and how the mechanisms that produce processes such as the FIP effect are generated.
Stellar observations enable us to explore a larger parameter space of stellar properties such as activity, rotation, 
spectral type etc. \citep{wood&linsky_2010}, whereas solar observations allow much more detailed analysis of specific 
features and scenarios.
It is from stellar observations that we now know an inverse FIP effect can happen, and that it occurs in later spectral
types with stronger magnetic fields, but high spatial and temporal resolution observations are not possible. 

Such observations were also not possible for the Sun until recently, when \citet{doschek_etal2015} discovered several
instances of abundance anomalies occurring near strong sunspot magnetic fields during solar flares. They found unexpectdly
high \ion{Ar}{14} 194.396\,\AA\, and 187.964\,\AA\, line intensities relative to \ion{Ca}{14} 193.874\,\AA. These lines are
formed in similar temperature conditions so the observed ratios were unusual. Their measurements are the first detection
of the inverse FIP effect on the Sun, and the first observations of an inverse FIP region with high spatial and temporal 
resolution. \citet{doschek&warren_2016} subsequently identified several further events. 
With the discovery of these events, we can bring the power of spatial and temporal resolution to bear on understanding
the inverse FIP effect. 

An important question, that is difficult to answer from stellar observations, is whether the
inverse FIP effect is caused by an enhancement of high FIP elements (HFE) or a depletion of low FIP elements (LFD). This
is significant for several reasons. 
A promising model to explain the FIP and inverse FIP effects has been developed by 
\citet{laming_2004, laming_2012} based on the ponderomotive force acting on Alfv\'{e}n waves in coronal loops. In this model,
the FIP or inverse FIP effect arises from the direction of propagation of the waves. Waves with a coronal origin - perhaps
excited by nanoflares - cause the usual FIP effect by enhancing low FIP elements in the corona. Waves with a sub-photospheric
origin, that might be prevalent around sunspots, cause the inverse FIP effect by depleting low FIP elements in the corona.
In both cases, there is a clear prediction that the ponderomotive forces are acting on low FIP elements. 
Furthermore, recently a solar cyclic variation in coronal elemental abundances has been found
for the Sun when it is observed as a star \citep{brooks_etal2017}. This has important implications for comparisons of coronal
abundances with fixed stellar properties, but the magnitude of the effect is not fully understood in stars. 

For the Sun,
there is some evidence that the cyclic variation results from changes in the low FIP elements, since a similar cyclic variation is not seen 
directly in the high FIP element Ne \citep{schmelz_etal2005, delzanna&andretta_2011}. On the contrary, the Ne/O abundance ratio 
appears to vary with the solar cycle in the solar corona and solar wind \citep{shearer_etal2014, landi&testa_2015, brooks_etal2018},
which could in principle be due to changes in the Ne abundance. 
Solar-like stars of a similar spectral type 
to the Sun may show cyclic variations, but it may be more difficult to detect on later spectral type stars - where the inverse FIP effect is
seen. These stars are fully convective and may be less likely to show cyclic variations anyway, but confirmation
of the nature of the effect may further implicate sub-photospheric drivers as the cause of the inverse FIP effect. 

\citet{doschek_etal2015} tried to determine which elements were being affected in their observations by looking at the 
spatial homogeneity of the \ion{Ar}{14} and \ion{Ca}{14} emission, but did not draw a firm conclusion. \citet{doschek&warren_2016}
followed a method originally proposed by \citet{delzanna_2013b} to address the same question. In this method, the path length is
calculated and compared directly to the loop spatial dimensions (length and width). If the path length is larger/smaller than expected
from the observational measurements then it implies that something is over/under-estimated in the calculation. Since the intensity,
electron density, and line contribution function are all measured or assumed known, the remaining factor in the calculation
is the elemental abundance. \citet{doschek&warren_2016} concluded that the low FIP elements were depleted in the case they analyzed
because their calculated path length for the low FIP element Ca was too small, suggesting that its abundance was depleted.
The conclusion relies on the \citet{delzanna_2013b} method, which assumes a filling factor of one. This may (or may not)
apply to particular post-flare loops; see for example \citet{teriaca_etal2006}.

The first event found by \citet{doschek_etal2015} is interesting (flare of 2014, December 20), because it 
appears to result from the simultaneous release of magnetic energy across a post-flare loop arcade, with many of the
loops exhibiting similar properties in terms of width, length, temperature, density, and evolution. The cooling time
for loops with similar properties is broadly the same, and \citet{winebarger_etal2003a} showed that the expected lifetime
of loops at each temperature is also related to the observed delay between the peak emission at different wavelengths. 
Under the assumption that the loop cooling time is dominated by radiative cooling, \citet{aschwanden_etal2003} also pointed out
that the time delay between the peak emission at different temperatures can be related to the radiative loss function
and the low FIP element abundance enhancement factor. In the case where low FIP elements are depleted, the cooling delay
is longer. 
Motivated to search for a clear diagnostic of HFE or LFD we explore the 
impact on the radiative cooling of similar loops with different radiative power loss functions since
the most significant
difference between the post-flare loops and the inverse FIP region, in this flare, would appear to be the coronal elemental
composition itself. 
As a result, we develop
a new diagnostic that can distinguish between the two scenarios. We then apply the diagnostic in a preliminary analysis
of the 2014, December 20, inverse FIP event. 
\section{modeling}
\label{modeling}

\begin{figure}
  \centerline{\includegraphics[width=0.95\linewidth]{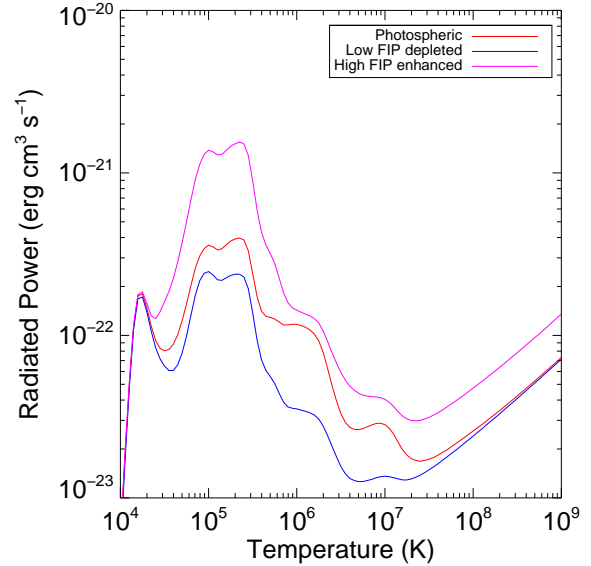}}
  \vspace{-0.1in}
\caption{Radiated power loss as a function of electron temperature calculated at an electron
density of 10$^{10}$ cm$^{-3}$ for three cases: photospheric abundances (red), a depletion of
low FIP elements (blue), an enhancement of high FIP elements (magenta). The functions calculated
with low FIP element depletion and high FIP element enhancement show opposite behavior with 
respect to the photospheric radiated power loss. The high continuum emission above 10\,MK
in the HFE model is due to the enhancement of He. 
\label{fig:fig1}}
\end{figure}

The total radiated power emitted from an optically thin plasma can be expressed (in units of erg 
cm$^{-3}$ s$^{-1}$) as
\begin{equation}
E_R = N_e N_H \Omega (T_e, N_e)
\end{equation}
where $N_e$ is the electron density, $N_H$ is the hydrogen density, $T_e$ is the electron temperature
and $\Omega (T_e, N_e)$ is the radiative power loss function. 
To assess the impact of the inverse FIP effect on the radiative cooling of loops we computed
$\Omega (T_e, N_e)$ for three cases. Recent studies of elemental abundances 
in solar flares using the most up-to-date atomic data, and comprehensive observations 
from SDO/EVE, show that flares predominantly evaporate photospheric plasma \citep{warren_2014}. 
So we adopted the photospheric abundances of \citet{grevesse_etal2007} as a base-line for 
``normal" radiative cooling in post-flare loops. The calculation was made using the CHIANTI v8
atomic database \citep{dere_etal1997, delzanna_etal2015} adopting the ionization fractions from
\citet{dere_etal2009}. The curve is shown in Figure \ref{fig:fig1}.

For the inverse FIP effect, we made two further calculations. To model the effect of low FIP 
element depletion (LFD model), we adopted a set of abundances computed as follows. We first calculated the
ratio of the photospheric abundances of \citet{grevesse_etal2007} to the coronal abundances of \citet{feldman_etal1992}.
The original photospheric
abundances of \citet{grevesse_etal2007} were then multiplied by this ratio, or depletion factor. 
The method ensures that all elements that are not
enhanced in the corona (high FIP) maintain abundances close to their photospheric values in the inverse model.
In contrast, low FIP elements are depleted in the inverse model by a factor comparable 
to the usual degree of coronal enhancement.
To model the effect of high FIP element enhancement (HFE model), we took
the average enhancement factor ($\sim4$) for the low FIP elements in the coronal abundances of 
\citet{feldman_etal1992}, and applied it to the high FIP element abundances only. In the HFE
model, the low FIP elements retain their photospheric abundances. Note that
for elements with FIP lower than 10\,eV\, the enhancement factors in \citet{feldman_etal1992}
fall in the range 3.9--5.7, so taking the average is close to the lower end of this scale. 
A higher value would lead to greater radiative losses and therefore faster cooling. Conversely,
we could have made a different choice of abundances for both the LFD and HFE models. Values
in the literature show a range of average enhancement factors and adopting lower values would clearly 
push the LFD and HFE radiative loss curves back towards the photospheric case. Comparing, for example,
the photospheric abundances of \citet{caffau_etal2011} with the coronal abundances of \citet{schmelz_etal2012},
we find only a factor of 2 enhancement.
All the other atomic data
used in the LFD and HFE models are the same as in the photospheric abundances case. 
The LFD and HFE model radiative power loss functions are also shown in Figure \ref{fig:fig1}. 
Machine readable text files containing these functions are available in the online version
of the paper. 

Figure \ref{fig:fig1} shows very significant differences between the three models. The peak of the
radiative power loss in the LFD model is about 60\% of the peak in the case of photospheric 
abundances. In contrast, the peak of the radiative power loss in the HFE model is about a factor
of 4 higher than in the photospheric abundances case. What is most interesting is that the LFD and 
HFE models show opposite behavior with respect to the photospheric radiative loss function.

To explore the impact of these calculations on post-flare loop cooling in detail, we have 
performed a flare simulation, incorporating all three $\Omega (T_e, N_e)$ functions, 
using the Enthalpy-Based Thermal Evolution of Loops (EBTEL) 
hydrodynamic model \citep{klimchuk_etal2008, cargill_etal2012}. EBTEL solves simplified versions
of the hydrodynamic equations that treat field-aligned average values of temperature, pressure,
and density. The plasma response to an impulsive energy release calculated by EBTEL has been compared
to results from more sophisticated 1-D hydrodynamic simulations 
including the Palermo-Harvard code \citep{peres_etal1982}, the Naval Research Laboratory
solar flux-tube model \citep{mariska_etal1982}, and the Adaptively Refined Gordunov Solver \citep[ARGOS,][]{antiochos_etal1999}.
Good agreement was found \citep{klimchuk_etal2008}. For details of the assumptions and methods
used in designing EBTEL see the papers referenced above.

The use of EBTEL restricts our analysis to a very simplified model. Several of the assumptions
place limitations on our results and it is appropriate to comment on some of them here. First, 
there is no provision for specifying the location of the flare energy release or the geometrical structure 
of the loop. These can affect the response to the heating. In this work, however, our principle goal
is to investigate the diagnostic potential of the method by examining how the loop evolution changes when we use different $\Omega (T_e, N_e)$ functions, 
so restricting other parameters in the model is not an initial concern. Second, because the EBTEL simulation 
treats field aligned average values we have to use a spatially invariant 
radiative power loss function. This
is also an approximation since actual observations seem to hint that abundances, and therefore
radiative losses, can be different between the loop footpoints and the apex \citep{baker_etal2013}. 
In fact, in the event we analyze here, the inverse FIP effect appears confined to the footpoint regions of the
post-flare loop arcade. These observations suggest that the loop cooling could be different between footpoint
and apex, but our results assume that the loop is cooling as a whole. It would be interesting
to compare our results and these observations with more sophisticated hydrodynamic simulations incorporating spatially and temporally
dependent radiative cooling in the future. 

Note that EBTEL uses parameterized versions of the radiative loss function for computational speed.
Since we are not concerned with speed for our single specialized simulation, we modified the EBTEL
procedures to input our calculated radiative power loss functions at the full temperature resolution.
For the interested reader, we found that using the full temperature resolution results in a factor
of 3.4 slow down in calculation speed. This is not important for our study, but may be significant for
simulations involving large numbers of field lines. 

We simulated the flare as a single impulsive heating event on a single thread. Properties of the loop in
the model of course affect the simulation. For example, the loop length affects the loop lifetime. We chose the loop properties
to broadly agree with those of the post-flare loop arcade in the inverse FIP flare (see section \ref{observations}). 
The loop half-length, $L$,
is 36\,Mm, and the loop radius, $\sigma$, is 775\,km. 
The heating is in the form a background static rate of 5$\times$10$^{-6}$ erg cm$^{-3}$ s$^{-1}$ and then
the imposition of an impulsive heating rate 4000 times stronger than the background i.e. 2$\times$10$^{-2}$ erg cm$^{-3}$ s$^{-1}$.
The heating event is a step function that is 
switched on for 300\,s and was chosen to raise the loop temperature to 
19.5--21.5\,MK. The density exceeds $\log N_e$ = 9.5 in all three 
cases. The loop then drains and cools.

\begin{figure*}
  \centerline{\includegraphics[width=0.95\textwidth]{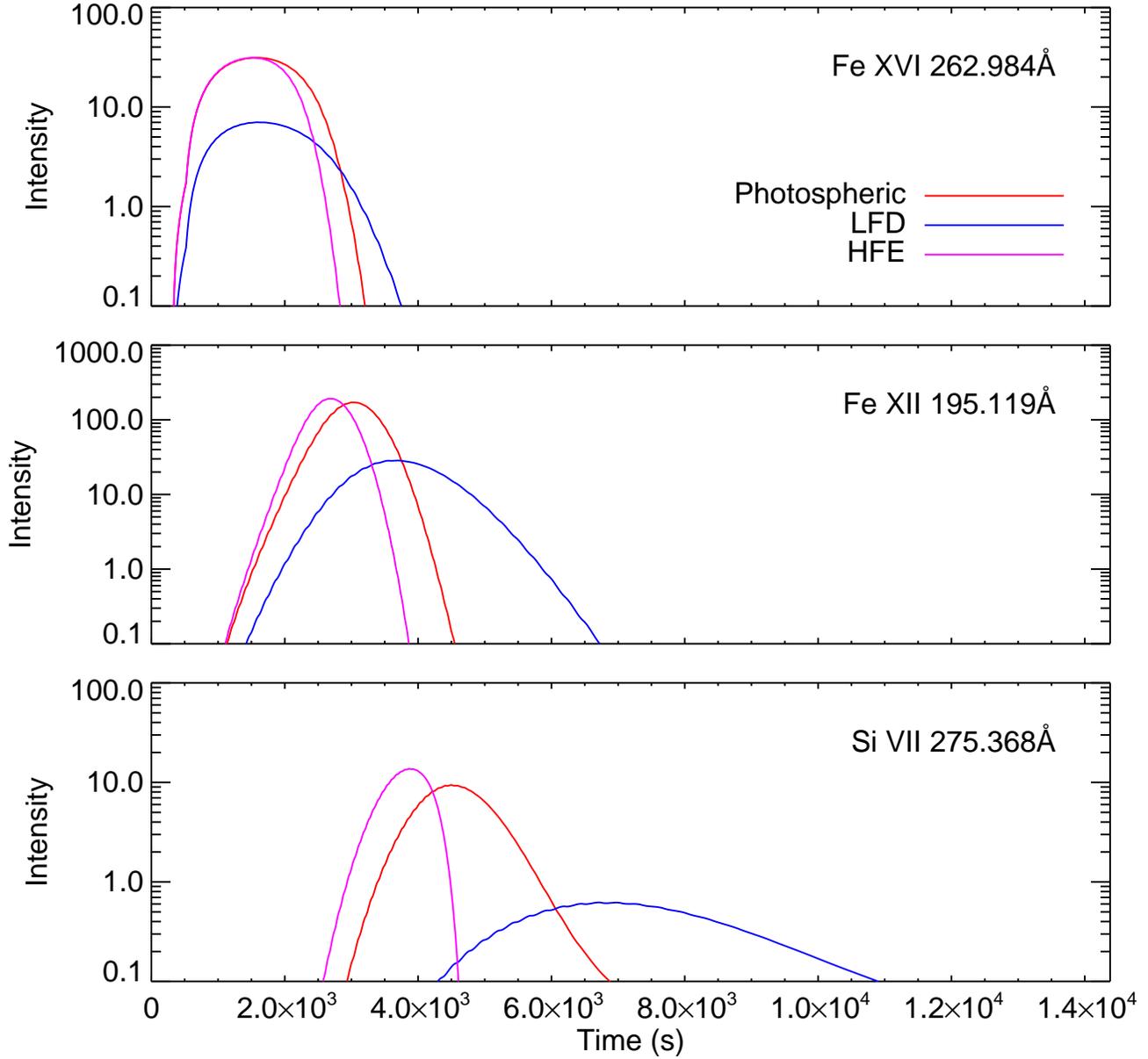}}
  \vspace{-0.1in}
\caption{Computed intensities of the \ion{Si}{4} 275.368\,\AA, \ion{Fe}{12} 195.119\,\AA, and \ion{Fe}{16} 262.984\,\AA\,
spectral lines based on an EBTEL simulation of a single impulsive heating event in a single strand. The simulation was
performed for three different radiated power loss functions, representing photospheric abundances (red) and two inverse
FIP models: low FIP element depletion (LFD, blue) and high FIP element enhancement (HFE, magenta). There are increasing 
differences in the strand lifetime going from the higher temperature line (\ion{Fe}{16} 262.984\,\AA)
to the lower temperature line (\ion{Si}{4} 275.368\,\AA). 
\label{fig:fig2}}
\end{figure*}

We simulated the spectral emission for three lines covering a broad range of temperatures using the flare 
simulations based on 
each abundance model for $\Omega (T_e, N_e)$. 
We used \ion{Si}{7} 275.368\,\AA, \ion{Fe}{12} 195.119\,\AA, and \ion{Fe}{16} 262.984\,\AA.
These lines are formed at 0.63\,MK, 1.58\,MK, and 2.75\,MK, respectively. We calculate the spectral line intensities using the formula
\begin{equation}
I_{ij} = A (Z)\, G (T, N)\, N^2\, ds
\end{equation}
where $I_{ij}$ is the intensity arising from a transition from atomic level $i$ to $j$, $A (Z)$ is the
adopted elemental abundance, $G (T, N)$ is the contribution function (containing all of
the - assumed known - atomic physics of the line formation, including in this definition the hydrogen to electron density ratio), 
and $ds$ is the column depth. $N$ and $T$ are the field-aligned average values of $N_e$ and $T_e$ output from the EBTEL 
simulation, and $G (T, N)$ is interpolated to these values. It therefore represents the emission averaged over the whole loop.

There is a complex relationship between measured
loop width and actual loop radius \citep{lopezfuentes_etal2006, brooks_etal2012}. Here we assume that 
the measured width is representative of the loop radius, and 
define $ds$ as 2$\sigma$. For computing the emission measure, $N_e^2 ds$, the electron density
is by far the dominant term, so this assumption does not greatly affect the computed intensities.

The simulation assumes that ionization equilibrium has had time to establish. This is not an unreasonable 
assumption for this event. \citet{doschek_etal2015} measured densities in the range $\log N_e$ = 10.5--11.5
for the inverse FIP region. At the lower density, the ionization relaxation timescales for Si and Fe at the
formation temperatures of \ion{Si}{7} 275.368\,\AA, \ion{Fe}{12} 195.119\,\AA, and \ion{Fe}{16} 262.984\,\AA\,
are less than 30\,s in a constant density model \citep{lanzafame_etal2002}, whereas emission in the highest 
temperature \ion{Fe}{16} 262.984\,\AA\, line does not form in our simulation until after 11\,mins. We define
the line formation time as the time at which the emission reaches 25\% of the peak.

\begin{figure*}
  \centerline{\includegraphics[width=0.95\textwidth]{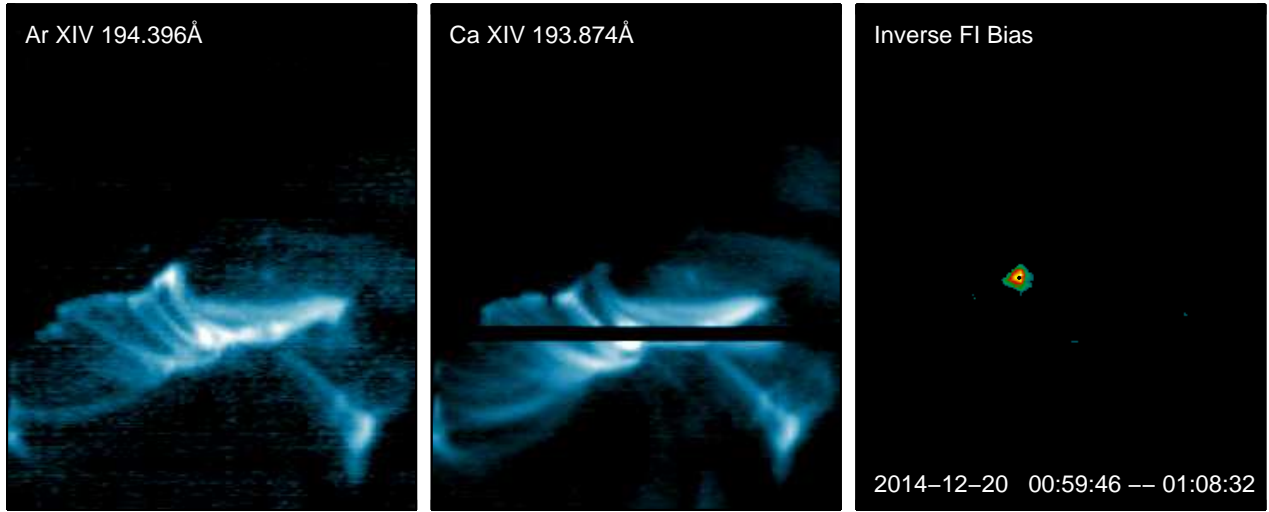}}
  \vspace{-0.1in}
\caption{EIS observations of the post-flare loop arcade following the X1.8 event on 2014, December 20, and
the detection of the inverse FIP effect. {\it Left panel: } \ion{Ar}{14} 194.396\,\AA\, slit raster
scan image. {\it Center panel: } \ion{Ca}{14} 193.874\,\AA\, image. {\it Right Panel: }
\ion{Ar}{14} 194.396/\ion{Ca}{14} 193.874 line intensity ratio. The ratio image was filtered to reduce
noise and only retain significant signals (see text). It was also normalized to the photospheric abundance
ratio.
\label{fig:fig3}}
\end{figure*}

We computed the contribution functions at each timestep using the simulation densities and temperatures and
the photospheric abundances of \citet{grevesse_etal2007} adjusted by the same factors used to construct the 
radiative loss functions in the HFE and LFD models. We considered photospheric abundances for comparison with 
the majority of the normal post-flare loops in the inverse FIP flare. The method we develop,
however, is not sensitive to the abundances
used for the contribution functions. The magnitude of the intensities will increase or decrease, but the
lifetime of the cooling loop is unchanged, and this is the key diagnostic. 

\begin{figure*}
  \centerline{\includegraphics[width=0.95\linewidth]{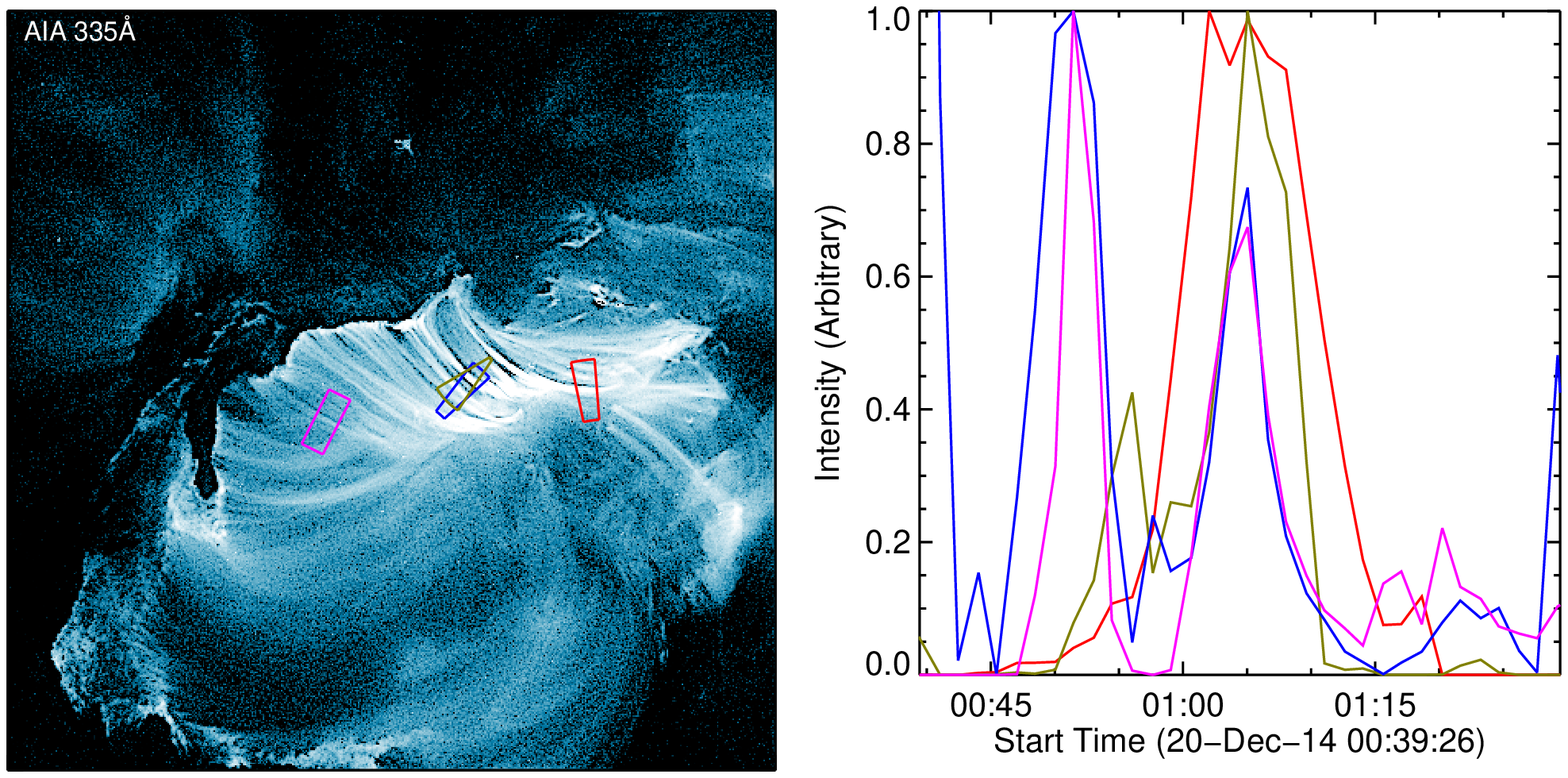}}
  \centerline{\includegraphics[width=0.95\linewidth]{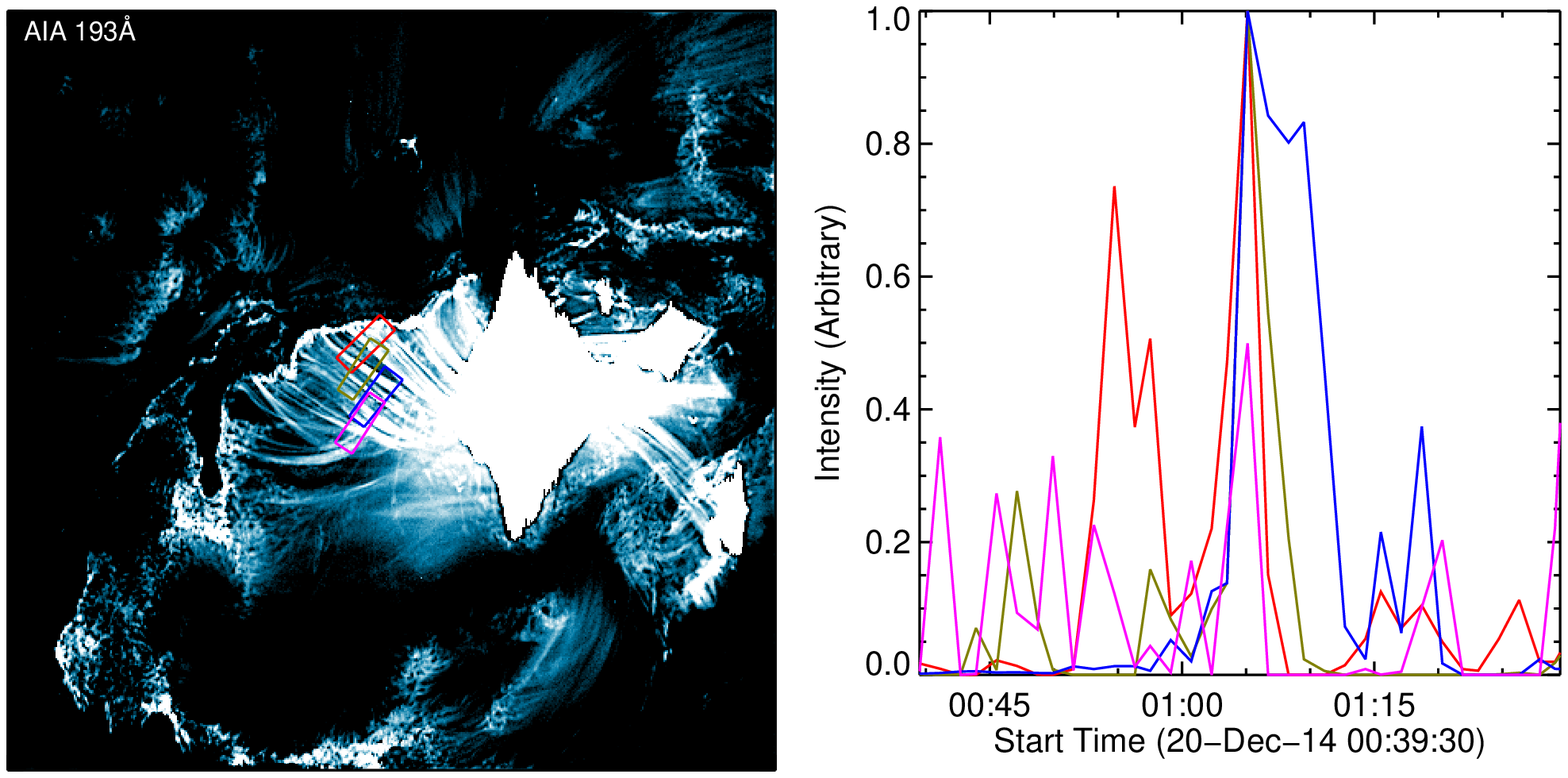}}
  \vspace{-0.1in}
\caption{AIA observations of the post-flare loop arcade and the behavior of a sample of loops. {\it Left panels: }
335\,\AA\, image taken at 1:05:02UT (top) and 193\,\AA\, image taken at 1:05:06UT (bottom). The colored boxes show the
loop segments examined for analysis and used to produce the light curves in the right hand panels. {\it Right panels: }
Light curves for the sample of loop segments (335\,\AA\, - top; 193\,\AA\, - bottom) around the time of the
inverse FIP detection. The different colors represent the different loop segments and the colors are matched.
The lifetimes of the post-flare loops are short at both wavelength: less than 17 mins in 335\,\AA\,
and less than 10 mins in 193\,\AA.
\label{fig:fig4}}
\end{figure*}

We show the results in Figure \ref{fig:fig2}. As we already hinted, the calculations based on the LFD and
HFE models show opposite behavior with respect to the computations based on photospheric abundances. Defining the
lifetime as being the time-period when the intensity is above 25\% of the maximum, the higher temperature (\ion{Fe}{16})
emission lasts longer in the LFD model (2280\,s) than the photospheric case (1920\,s), but the emission 
from the HFE model is shorter (1680\,s); though not very different. This difference is accentuated at lower temperatures. 
At the formation temperature of \ion{Fe}{12}, 1.58\,MK, the emission from the HFE model again has a lifetime slightly
shorter, but close to, that of the photospheric abundance case; 1020\,s and 1280\,s, respectively. Conversely, the 
lifetime for the LFD case is much longer (almost double): 2380\,s. The differences are even more dramatic at the 
formation temperature of \ion{Si}{7} (0.63\,MK). While the lifetime of the loop in the HFE model is 1140\,s compared
to 1840\,s for the photospheric abundance case, the loop persists 5 times longer (5540\,s) in the LFD model.

In the introduction we mentioned that the time delay between the peak emission at different temperatures is affected by
the radiative loss function. The cooling delay is longer when low FIP elements are depleted. 
We can also see this in Figure \ref{fig:fig2}. The time delay between the peak emission in \ion{Fe}{16} and
\ion{Fe}{12} is 1400\,s when photospheric abundances are assumed, but is shorter (1120\,s) and longer (2120\,s) in the HFE and
LFD models, respectively. Again, the difference is larger at lower temperatures. The time delay between the peak
emission in \ion{Fe}{12} and \ion{Si}{7} is 1480\,s using photospheric abundances, 1220\,s in the HFE model, and 3000\,s in the LFD model.

To summarize, for loops with similar properties, the radiative cooling due to a depletion of low FIP elements results
in a significantly extended emission lifetime compared to that of normal post-flare loops. The time delays between emission at 
different temperatures are also longer. An enhancement of high FIP elements
produces the opposite effect.
These differences are potentially detectable in inverse FIP observations. 
\section{observations}
\label{observations}

As a demonstration of the practical usage of our new diagnostic, we examine the flare of 2014, December 20, 
where the first evidence of the inverse FIP effect on the Sun was discovered by \citet{doschek_etal2015}.
The flare under discussion was an X1.8 event. It began around 00:10UT and the GOES X-ray flux peaked at
00:28UT. An extensive post-flare loop arcade was formed, with loops persisting well into the long duration 
decay phase of the event. The X-ray flux did not fall below M-class until after 03:00UT. 

The inverse FIP effect was seen due to enhanced \ion{Ar}{14} 194.396\,\AA\, emission, relative to \ion{Ca}{14} 193.874\,\AA, 
from around 00:15UT and was strongest around 01:00UT. Figure \ref{fig:fig3} shows raster scan data in the two spectral lines 
from the Extreme ultraviolet Imaging Spectrometer \citep[EIS,][]{culhane_etal2007a}
on {\it Hinode} \citep{kosugi_etal2007}. The data were obtained in flare observation mode by responding to 
the trigger from the X-ray Telescope \citep[XRT,][]{golub_etal2007}. Once triggered, the 2$''$ slit was
deployed to scan a field-of-view (FOV) of 240$''$ by 304$''$ in coarse 3$''$ steps. The exposure time was 5\,s so
that the FOV was covered in just under 9\,mins. An extensive linelist was telemetred to the ground. 
The flare response stopped running around 01:10UT.

We show a measure of the FIP bias (ratio of coronal to photospheric abundance) in the right hand panel
of Figure \ref{fig:fig3} which is constructed from the \ion{Ar}{14} 194.396\,\AA\, to \ion{Ca}{14} 193.874\,\AA\,
line intensity ratio. The image was filtered at 2.5\% of the peak intensity in both lines to reduce noise,
and only signals above the calibration uncertainty in the intensity ratio were retained. The ratio was 
also normalized to the photospheric abundance ratio. The figure clearly shows the location of the 
strongest inverse FIP signal. 

The EIS data were processed using standard procedures available in {\it SolarSoft}
i.e. EIS\_PREP. We applied the ground radiometric calibration \citep{lang_etal2006} since the two lines
are very close in wavelength and the degradation ratio between them shows a difference of less than 3\% according
to both \citet{delzanna_2013a} and \citet{warren_etal2014}. In any case, 
we only use the FIP bias to locate the inverse FIP region and do not make
use of any quantitative data. 
We collected data from the Atmospheric Imaging Assembly \citep[AIA,][]{lemen_etal2012} on the Solar Dynamics
Observatory \citep[SDO,][]{boerner_etal2012} covering 2 hours before and after the strongest inverse FIP signal
is seen. These data are level-1 and have been flat-fielded and normalized to the 2.9\,s exposure time. 

To locate the inverse FIP region on the AIA images we coaligned a 335\,\AA\, filter image taken at 01:05UT 
with the EIS \ion{Ar}{14} 194.396\,\AA\, raster scan taken at 00:59UT. The 335\,\AA\, filter has a strong
peak at 2.5\,MK\, resulting from \ion{Fe}{16} emission so is closest in formation temperature to the \ion{Ar}{14}
line (3.4\,MK). The AIA image was re-scaled to the EIS pixel grid size and the two images were coaligned
through multiple cross-correlation steps. 

\begin{figure*}
  \centerline{\includegraphics[width=0.95\textwidth]{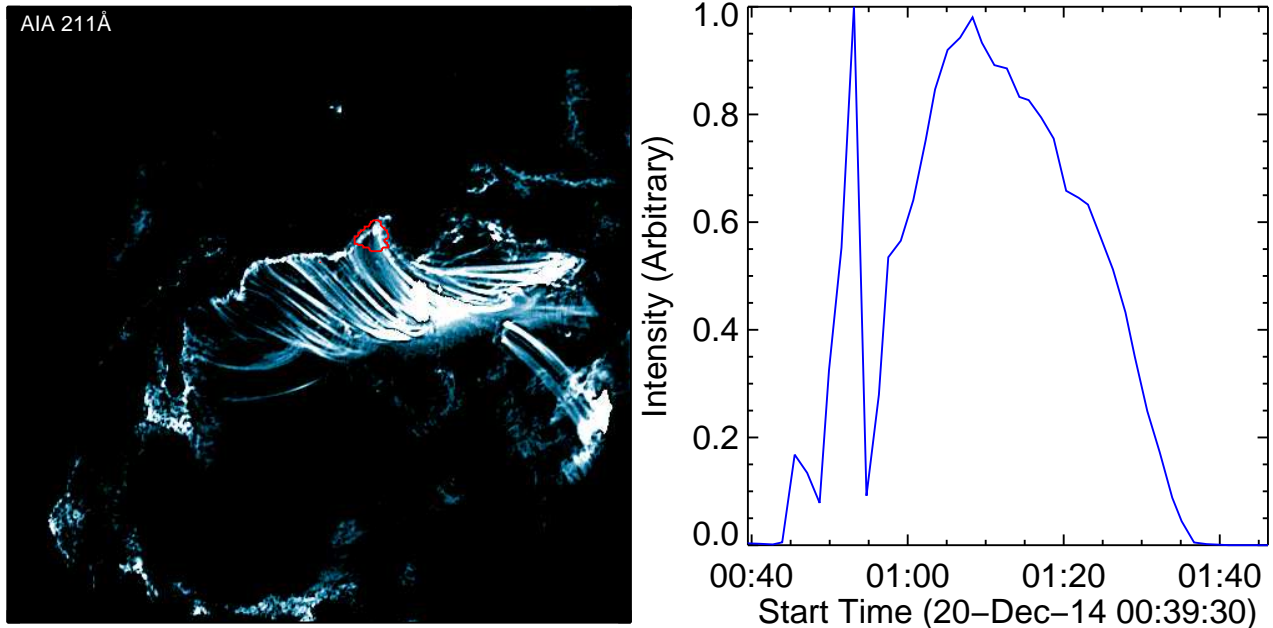}}
  \vspace{-0.1in}
\caption{{\it Left panel: } AIA 211\,\AA\, image of the post-flare loop arcade showing the location of the inverse
FIP region
from Figure \ref{fig:fig1} (red contour). The image was taken at 1:05:11UT. {\it Right panel: } AIA 193\,\AA\, intensity
evolution of the inverse FIP region. The lifetime of the inverse FIP region ($\sim$ 42 min) is much longer than for the post-flare
loops in Figure \ref{fig:fig4}.
\label{fig:fig5}}
\end{figure*}

We show examples of the behavior of the post-flare loop system in the AIA 335\,\AA\, and 193\,\AA\, filters
in Figure \ref{fig:fig4}. We picked out a few (ten) loops in the arcade for spot checks and measured their lengths, 
widths, and lifetimes. To do this, we followed the procedure of \citet{aschwanden_etal2008} (as modified and 
implemented by \citet{warren_etal2008a}), and identified relatively isolated loops, selected a clean segment, straightened
them, and averaged the cross-loop intensity profiles along the segments. 
We then selected two positions at the edges of the loops to identify the background emission, and 
fit a Gaussian function to the averaged background subtracted cross-loop intensity profile to 
obtain the loop width (Gaussian
width, $\sigma$), and the total intensity. This procedure was then repeated throughout the time-series of
the observations to obtain the lightcurves. Furthermore, we measured the loop half-lengths, using the same
procedure, by selecting and straightening loop segments extending from the (visually identified)
footpoints to the loops' apex. The widths fall in the range 635--1280\,km (FWHM) and the half-lengths fall
in the range 17.9--35.8\,Mm. These measurements motivated the model parameters for our simulation in 
section \ref{modeling}.

The colored boxes in Figure \ref{fig:fig4}
indicate the segments we chose for the lightcurve (and loop width) measurements, and are different than the larger
segments we used for the half-length measurements.
Here we discuss the cleanest lightcurves for only four of the loops. 
The lightcurves were somewhat messy due to the fact that the loops are rapidly varying and moving in the field
of view, and because the flare saturates some frames of the 193\,\AA\, images, as we can see in Figure \ref{fig:fig4},
hence the reason for focusing on a restricted sample. In fact, as we can see in the Figure, several lightcurves
appear to exhibit multiple brightenings, either due to re-brightening of the same loop (see e.g. the blue curve 
in the top panel of Figure \ref{fig:fig4}), or due to the appearance of different loops (e.g. the
red curve in the bottom panel).

As in section \ref{modeling}, we define the loop lifetime as the time-period when the intensity is above 25\% of 
the maximum. Thus defined, the 335\,\AA\,
loop lifetimes fall in the range 6--17\,mins (360--1020\,s). 
In the 193\,\AA\, filter the post-flare loop lifetimes fall in the range 1.5--10 mins (90--600\,s).
We also checked that the lifetime of these loops that appeared around the time of the inverse FIP signal are 
representative of most of the loops in the post-flare arcade, by examining a sample of 20 loops that occurred 
15 and 5 mins before, and 5 and 15 mins after, the appearance of the IFIP region. Half of these loops had lifetimes
that fall within the same range, and all but one had a lifetime less than 17 mins (similar to the 335\,\AA\,
loops).
We conclude then that the post-flare loops in the arcade evolve on fairly similar timescales at each temperature, 
which is consistent with 
their broadly similar properties. We discuss the one exceptional 193\,\AA\, loop further below.

The observed loops are
somewhat more transient than found in our simulations, which probably reflects differences in the modeled and
observed properties. As noted, \citet{doschek_etal2015} measured densities in the range $\log N_e$ = 10.5--11.5
for the inverse FIP region,
whereas $\log N_e >$ 10.0 is difficult to achieve in our simulations. Higher densities would lead to stronger
radiative cooling and therefore shorter lifetimes.

Figure \ref{fig:fig5} shows the behavior of the 193\,\AA\, light curve for the inverse FIP region. The 211\,\AA\,
image in the left panel shows the location of the inverse FIP region. It is close to the footpoint of one of
the post-flare loops that is rooted in the strong magnetic field of the sunspot. The lightcurve is dramatically 
different from any of those found for the (small) sample of loops we checked. The duration is about 42 mins
($>$ 2520\,s). 
This is more than a factor of four longer than the lifetimes of the other loop segments we measured
in the 193\,\AA\, filter, and immediately suggests that low FIP elements are depleted.
Furthermore, the lifetimes of all the post-flare loop segments in 335\,\AA\, (and most of the segments in 193\,\AA)
are shorter than in our simulations
for any of the radiative loss functions. Only two of the 193\,\AA\, loop segments have lifetimes that come close to
the simulated durations from the HFE model. In contrast, the lifetime of the inverse FIP region in the 193\,\AA\,
filter is quite close to the lifetime simulated for the LFD inverse model (2380\,s). Although this result provides
only weak support (since the simulation clearly does not capture the evolution of all the loops correctly), it 
also agrees with the conclusion that the low FIP elements are depleted. 

Note that we assume that the inverse FIP region persists as long as the intensity enhancement seen in the images.
The EIS data do not cover the full duration of the extracted light curve in Figure \ref{fig:fig5} so strictly
speaking we cannot confirm this. They are, however, consistent with the long duration since the inverse
FIP region was detected by EIS at 00:47UT and is still visible at 01:05UT just before the EIS observations stopped. 
So even in the EIS data the anomaly exists for at least 18--27\, mins and there are no more data during the
remaining $\sim$30\, mins of the intensity enhancement.

As discussed above, there was one exceptional 193\,\AA\, loop we measured with a long duration. This loop had a 
lifetime of about 42\, mins, which is the same as that of the IFIP region. As with that region, it was measured
close to the loop footpoint, and this suggests that it could be a candidate to show the same abundance anomaly
effect. Unfortunately, the EIS slit does not appear to cross this footpoint region when the loop forms, so we
could not verify this conjecture.

\section{Summary and discussion}

We have developed a new diagnostic that can determine whether high FIP elements are enhanced, or low
FIP elements are depleted, in inverse FIP effect regions on the Sun. The method relies on the markedly
different radiated power loss functions that result from modeling the degree of enhancement or depletion
of the high- or low-FIP elements. In numerical hydrodynamic simulations of impulsively heated loops,
these radiated power loss functions lead to significantly different loop cooling times depending on
the specific abundance pattern. In particular, loops forming around 1--2\,MK last significantly longer
than loops filled with photospheric plasma if the low FIP elements are depleted. They persist even
longer at cooler (0.6\,MK) temperatures. In contrast, when high FIP elements are enhanced, loops
forming around 1--2\,MK cool more rapidly than loops filled with photospheric plasma. 

We applied the new method to the analysis of the X1.8 flare that occurred on 2014, December 20. The post-flare
loops in the arcade produced by the energy release show broadly similar properties such as width, length, 
temperature, and density, so would be expected to evolve on similar timescales if they are filled with 
plasma evaporated from the photosphere as in many flares \citep{warren_2014}. In the inverse FIP region 
the composition is different, however, and according to our diagnostic may show a different evolution. 
We found that the small sample of loops we analyzed lasted for $<$ 10\,mins in the AIA 193\,\AA\, filter.
Conversely, the inverse FIP region persisted for about 40\,mins. This suggests that low FIP elements
are depleted in the inverse FIP region. 

Our analysis is really a demonstration that the method can work. 
Many loops appeared during the evolution of the post-flare arcade and we only sampled a few that
existed at the time of the inverse FIP event. We do not claim that our measurements are representative
of all the loops in the arcade throughout the full duration of the flare. Our objective here is to
introduce our new diagnostic method, and show that it can be applied to real inverse FIP observations.
Further detailed analysis will be needed to fully understand the 2014, December 20 event. For example, 
one important point to mention is that as the flare ribbons sweep across the sunspot in the active region
they appear to be partly stopped in the inverse FIP region 
by the strong magnetic field of a lightbridge. This could lead to a pile-up of energy input
there, that may extend the loop footpoint lifetime, and could also be involved in the generation of the 
inverse FIP effect itself. 

Furthermore, our modeling scenario is fairly simplistic. We simulated a single impulsive event in 
a single strand, but the loop lifetime could be related to the number of strands rather than the cooling 
time. 
Most of the loops in the post-flare loop arcade could 
be composed of a
few strands, whereas the loop rooted in the inverse FIP region could contain relatively more strands.
The behavior of different strands in the region of the flare ribbons over the sunspot could also be important. 
We have implicitly assumed that the inverse FIP region is within the post-flare loops and that a loop model
is applicable to it. It could be the case, however, that the inverse FIP emission is a result of interactions between
the post-flare loops and the sunspot magnetic field.

Nevertheless, the results presented here are in agreement with the model of the FIP and inverse FIP effects proposed
by \citet{laming_2004}. In that model, the ponderomotive forces acting on Alfv\'{e}n waves result in 
changes in the behavior of the low FIP elements only. In the inverse FIP effect, the low FIP elements
are depleted, resulting in longer loop cooling times, as we find here.

It is also interesting that the inverse FIP region is confined close to the loop footpoint and does not
fill the whole loop. This behaviour is reminiscent of observations of the normal FIP effect in active region loops.
\citet{baker_etal2013} show examples of loops with coronal composition near the footpoint, and traces
of enhanced composition along parts of the loops. Their argument is that these signatures are the first
signs of fractionated plasma mixing in the loops. If the FIP effect is caused by the ponderomotive
force as in the \citet{laming_2004} model, and the inverse effect is caused by the same mechanism only
acting on oppositely directed Alfv\'{e}n waves, then we might expect similar signatures of the process.
Unfortunately there are no EIS observations taken later in the flare to examine whether the post-flare loops become 
completely filled.

Our results may be of interest to stellar astronomers who observe both the FIP effect and inverse 
FIP effect in solar-like stars but have no comparable way to determine which elements are enhanced 
or depleted. The radiative power loss functions we provide could be used in hydrodynamic modeling
of the emission from these objects. They could also be used for simulations 
of abundance anomalies in flaring loops on active stars such as have been observed on the M dwarf 
CN Leonis or the eclipsing binary Algol: an evolution from sub-photospheric to photospheric
abundance was detected during giant flares on both \citep{liefke_etal2010,favata&schmitt_1999}.


\acknowledgments 

D.H.B. would like to thank Harry Warren and Deborah Baker for helpful comments on the manuscript. 
The work of D.H.B. was performed under contract to the Naval Research Laboratory and was funded
by the NASA \textit{Hinode} program. Authors are grateful to Konkoly Observatory, Budapest,
Hungary, for hosting two workshops on Elemental Composition in Solar and Stellar Atmospheres
(IFIPWS-1, 13-15 Feb, 2017 and IFIPWS-2, 27 Feb-1 Mar, 2018) and acknowledge the financial
support from the Hungarian Academy of Sciences under grant NKSZ 2018\_2. The workshops have
fostered collaboration by exploiting synergies in solar and stellar magnetic activity studies
and exchanging experience and knowledge in both research fields.
\textit{Hinode} is a Japanese mission developed and launched 
by ISAS/JAXA, collaborating with NAOJ as a domestic partner, NASA and STFC (UK) as international 
partners. Scientific operation of the \textit{Hinode} mission is conducted by the \textit{Hinode} 
science team organized at ISAS/JAXA. This team mainly consists of scientists from institutes in 
the partner countries. Support for the post-launch operation is provided by JAXA and NAOJ(Japan), 
STFC (U.K.), NASA, ESA, and NSC (Norway). CHIANTI is a collaborative project involving George Mason
University, the University of Michigan (USA) and the University of Cambridge (UK). Courtesy of
the NASA/SDO, and the AIA, EVE, and HMI science teams.


\end{document}